\documentclass[%
onecolumn,%
oneside,%
floats,%
aps,%
 prd,%
nobibnotes,%
nofootinbib,%
amsmath,%
amssymb,%
amsfonts,%
amscd,%
  superscriptaddress,%
eqsecnum%
]{revtex4}

\usepackage[utf8]{inputenc}
\usepackage{graphicx,array,dcolumn}
\usepackage{cases}
\usepackage[paperwidth=210mm,paperheight=297mm,centering,hmargin=1.8cm,vmargin=2.5cm]{geometry}
\usepackage{enumerate}

\usepackage{hyperref}
\hypersetup{hidelinks}

\newcommand{\bi}{\bibitem}

\def\be{\begin{eqnarray}}
\def\ee{\end{eqnarray}}

\def\-g{\sqrt{-g}}

\renewcommand\rho{\varrho}

\begin{document}

\title{
Solution of JWST and HST problems by primordial black holes
}

\author{A.D. Dolgov}
\email{dolgov@nsu.ru}
\affiliation{Novosibirsk State University, Novosibirsk, 630090, Russia}
\affiliation{Bogolyubov Laboratory of Theoretical Physics, JINR, Dubna, 141980, Russia}

\begin{abstract}

Observations of the early universe at redshifts of order 10, collected during the last several years presented by the Hubble Space 
Telescope (HST) and recent data by the James Webb Space telescope (JWST) created strong doubts on the validity of the conventional 
$\Lambda$CDM cosmology. It is argued here that the 30 year old conjecture by A.Dolgov \& J.Silk~\cite{DS} of galaxy and quasar 
seeding by supermassive primordial black holes (SMBHs) naturally solves the problem of the observed dense early population of the 
universe, as well as the problem of the formation of SMBH in the contemporary universe. The idea of seeding of cosmic structures by 
SMBH is supported  by a good agreement of the predicted log-normal mass spectrum of primordial black holes (PBH) with observations 
and by a noticeable antimatter population of the Milky Way. 
 
\end{abstract}

\maketitle

  \section{Introduction \label{s-intro}}

Recent observations of two orbital telescopes, HST and JWST created strong
doubts on the validity of the standard cosmology. However, the
30 year old conjecture~\cite{DS},  that the galaxy formation is
{\it seeded} by supermassive primordial black holes (PBH), not only rescued the 
accepted cosmological model but in fact predicted the observed rich population of the
early universe by well developed galaxies and quasars. 

The suggested mechanism of PBH formation allows for creation of sufficient density  
of massive PBHs that seeded structure formation not only in the early universe at redshifs of order ten but during
all cosmological epochs including the present day universe as well. Recently the idea of seeding by black holes
was rediscovered  in several papers under  the pressure of HST and JWST observations.

There is a strong confirmation of the model by "experiment".  Namely the
predicted log-normal mass spectrum of PBH very well agrees with observations;
noticeable antimatter population of the Galaxy is predicted and confirmed by
large amount of data: positrons, antinuclei, and (possibly) antistars, and  
the model could supply the sufficient number of SMBH binaries that presumably distorts the pulsar timing.

The paper is organised as follows. In section \ref{s-crisis} the observations of HST, JWST and ALMA are discussed and a
possible conflict of the data with the standard $\lambda$CDM cosmological model is described. In sec.~\ref{s-seeding} the 
seeding of cosmic structures by primordial black holes, originally proposed in our papers~\cite{DS,DKK} and recently 
rediscovered in several works, is discussed.  Next, in sec.~\ref{s-GW-BH}, the observations of gravitational waves generated 
by  the coalescing black hole binary systems is considered and the arguments in favour that these black holes are primordial
are presented. The chirp mass distribution are presented and compared with log-normal spectrum of PBHs. In sec.~\ref{s-black-DM}
the  possibility of PBH being the cosmological dark matte is advocated. Sec.~\ref{s-antimatter} is dedicated to discussion of 
antimatter in cosmology. In the following section observational data indicating to noticeable amount of antimatter  in the Milky Way is 
presented. Lastly in sec.~\ref{s-anti-creation} the mechanism of PBH creation with log-normal mass spectrum and antimatter formation
in the Milky way suggested in refs.~\cite{DS,DKK} is briefly described

\section{\label{s-crisis}
Crisis in cosmology}
\subsection{
Crisis in cosmology, is it real? \label{ss-crisis}}
Dense population of the early universe, younger than one billion years at redshifts
$z \sim 10$, discovered by Hubble Space Telescope (HST) and James Webb Space
Telescope (JWST), was taken as a strong blow to the conventional $\Lambda$CDM
cosmology.
However, the resolution of the problems by primordial black holes was
suggested long before these problems emerged, see refs.~\cite{DS,DKK} 
and several subsequent papers by our group.

The proposed model of the PBH creation is strongly supported by the predicted log-normal mass spectrum of the black holes: 
\be
\frac{dN}{dM} = \mu^2 \exp\left[ - \gamma \log^2 \left(M/M_0)\right)\right] ,
\label{log-normal}
\ee
where $\mu$ is an unknown normalisation parameter with dimension of mass, $\gamma$ is not determined by the model as well,
but the central mass $M_0$ is theoretically  predicted~\cite{AD-KP-M0} 
to be close  to
$M_0 \approx 10 M_\odot$, where $M_\odot \approx 2 \times 10^{33} $ gram is the solar mass.

\subsection{Comparison of JWST and HST \label{ss-comparioson}}

The orbit of HST is at the distance of 570 km from the Earth. The orbit of JWST is much larger, it is about $1.5 \times 10^6 $km. 
The mirror of HST has diameter equal to 2.4 m, while JWST has 2.7 time larger one and correspondingly the area of JWST mirror is 
approximately 7.4 times larger. In fig. 1 the images of HST and JWST are presented.
HST operates in optical wave length range, for example 450 nm, corresponding to blue light. It has also a possibility to catch the signal in the infrared 
range with the wave length 0.8-2.5 microns. JWST has high sensitivity to infrared radiation with the wave length 0.6 - 28,5 micron. It allows to penetrate 
deep into the early universe, up to redshifts $z \sim 15$.

\begin{figure}[htbp]
\begin{center}
\includegraphics[scale=0.30,angle=-90]{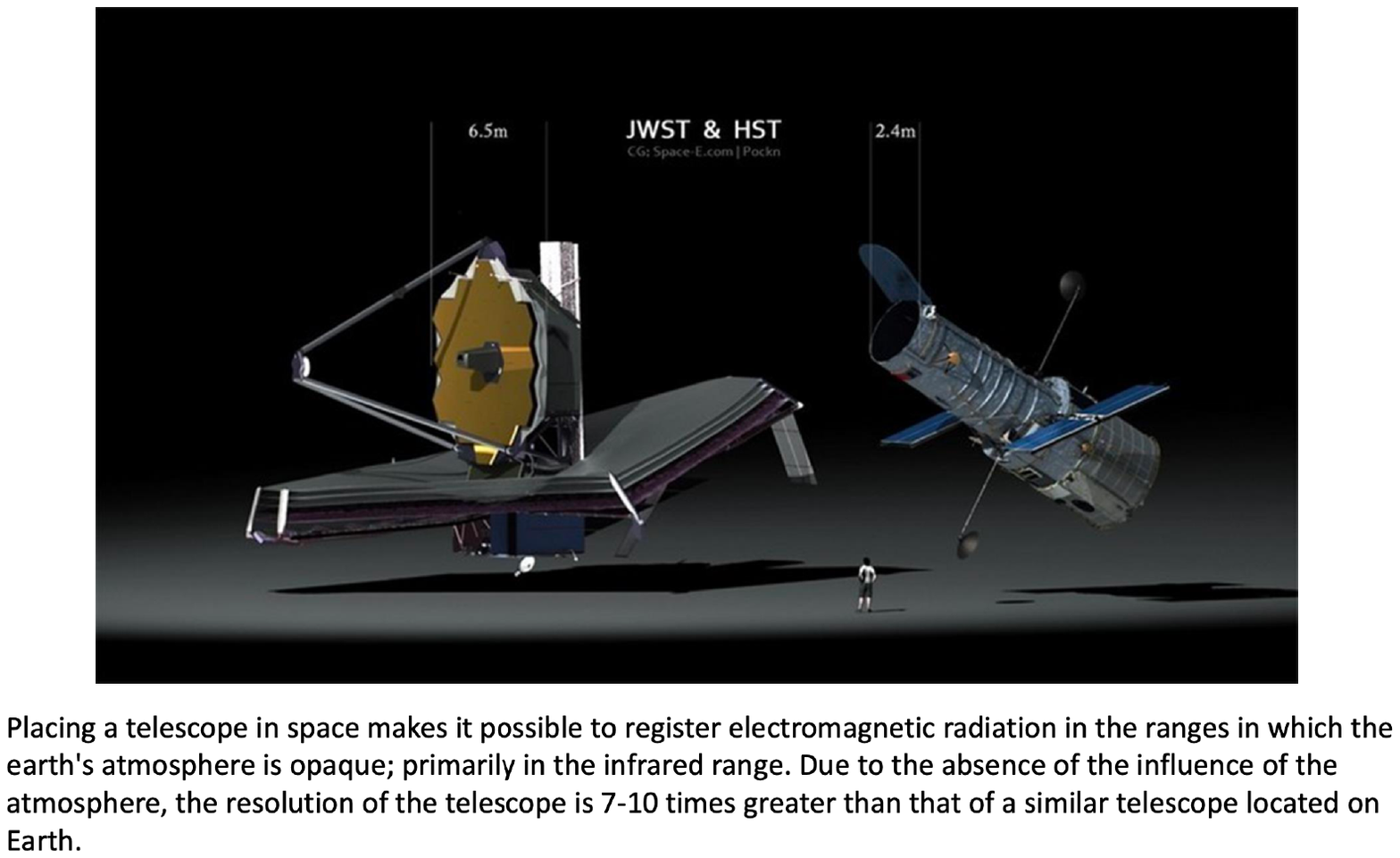}
\caption{Photos of HST and JWST }
\end{center}
\end{figure}

By a lucky chance of gravitational lensing HST was able to discover the galaxy at  $z \approx 12$ beyond the expected sensitivity.
The same galaxy was observed by JWST, as is depicted in  Fig. 2 .

\begin{figure}[htbp]
\begin{center}
\includegraphics[scale=0.28,angle=-90]{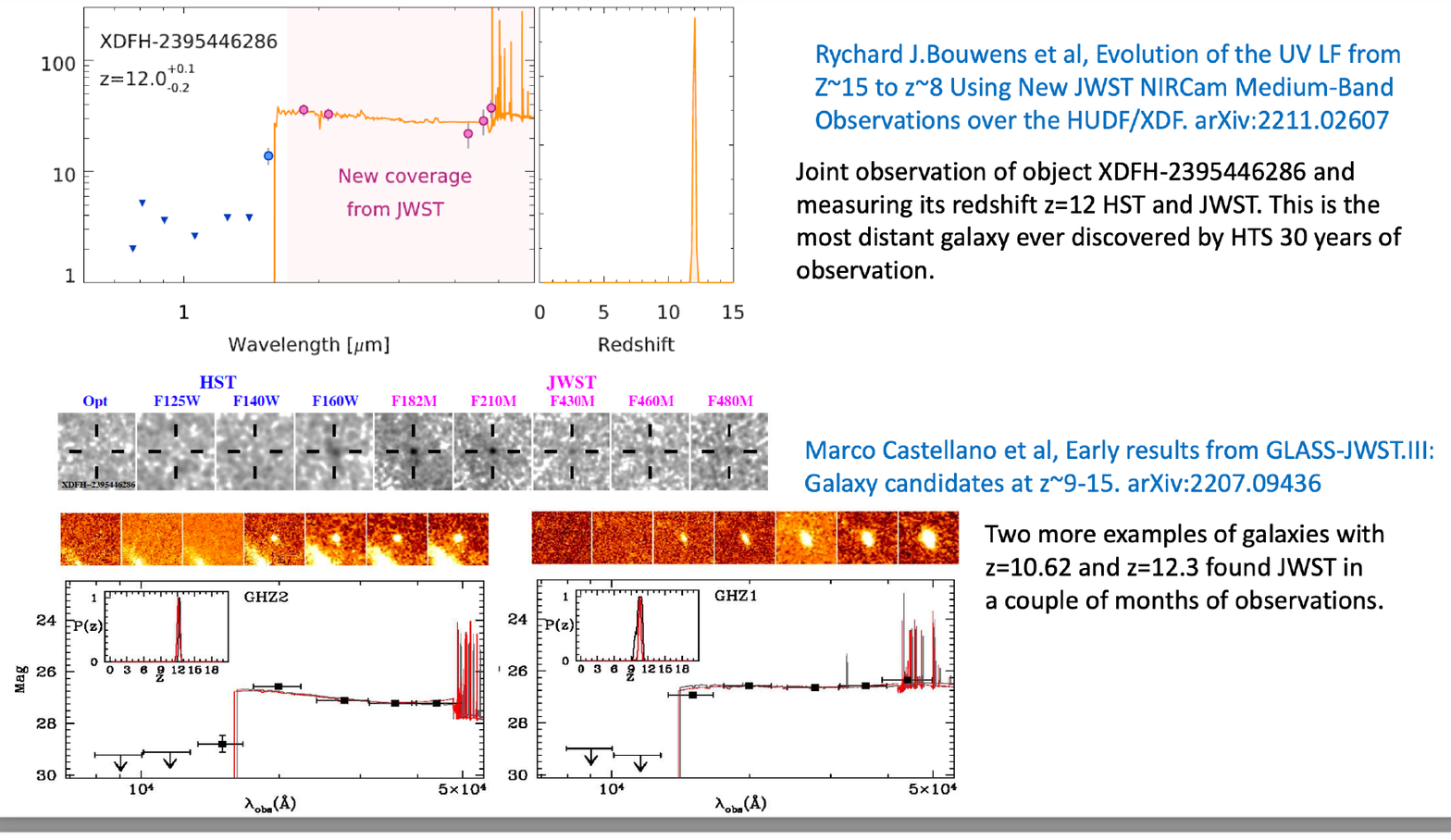}
\caption{The same galaxy observed by HST and JWST}
\end{center}
\end{figure}

\subsection{Early objects observed by JWST  and ALMA \label{ss-JWST-obs}}

The observations of far away galaxies discovered by JWST  at several high redshifts are  presented  in Fig.~3 and
compared with the theoretical expectations of the $\Lambda$CDM cosmological model. The difference  between the canonical 
theory and the data differs by several orders of magnitude especially  at $z \approx 15$.

\begin{figure}[htbp]
\begin{center}
\includegraphics[scale=0.28,angle=-90]{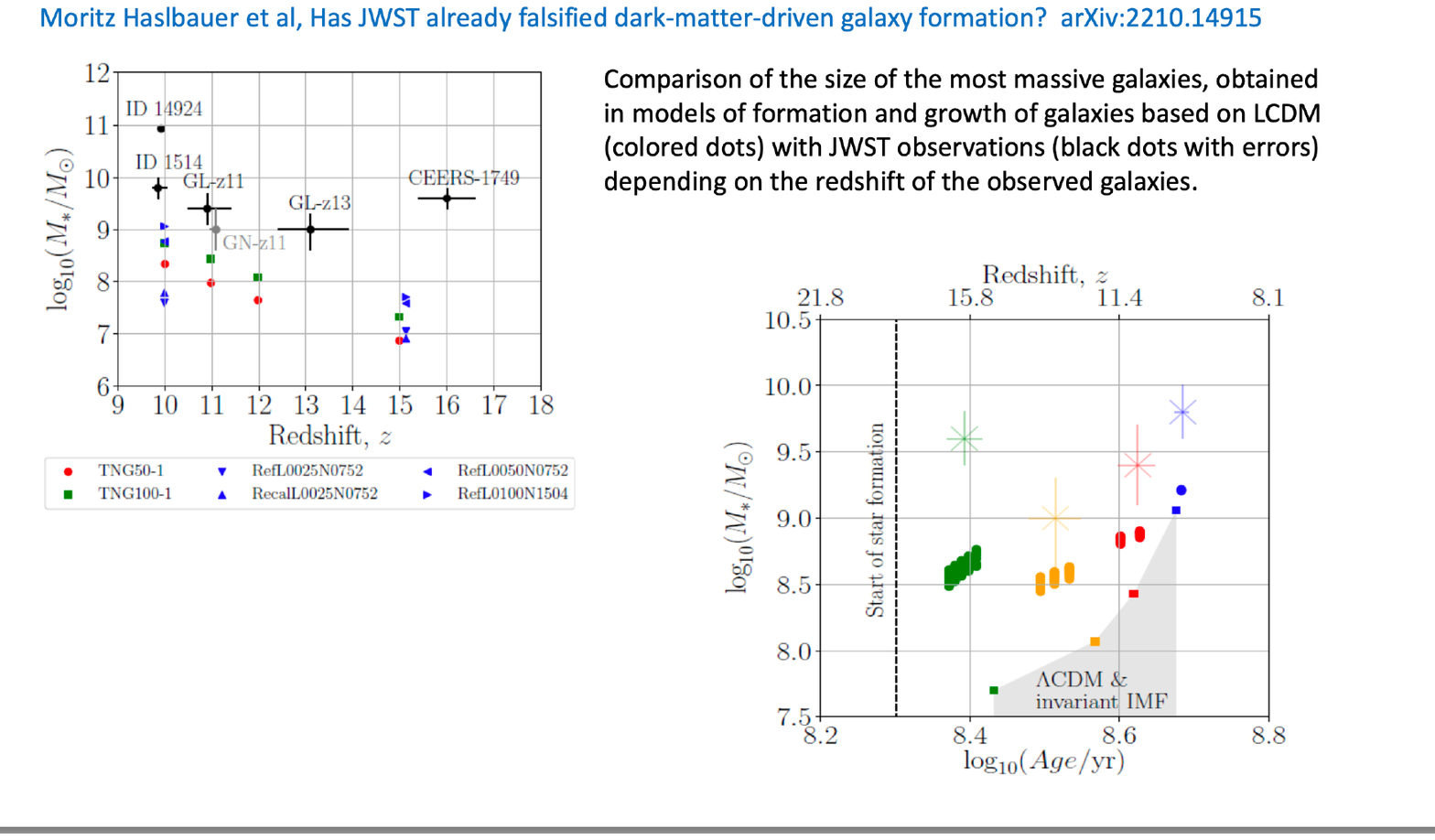}
\caption{JWST registration of high $z$ galaxies versus the predicted density from the $\Lambda$CDM model}
\end{center}
\end{figure}

Initially the discoveries of JWST  have been met with a grain of pepper, since
only continuum in micron range was measured till February 2023. That raised doubts
about accuracy of the determination  the observed galaxies redshifts. 

Quite soon, however, numerous spectral determinations presented excellent support of   the early data.

In ref.~\cite{Bunker} the galaxy GN-z11 has been studied and the spectroscopy confirms that GN-z11 is a 
remarkable galaxy with extreme properties seen 430 Myr after the Big Bang. The galaxy redshift is derived,
equal to z = 10.603 (somewhat lower than previous determinations) based on multiple emission lines in the low
and medium resolution spectra over $(0.8 - 5.3) \mu$m.

Another example of spectrum determination is presented by a different instrument,  the radio telescope array ALMA (Atacama
Large Milimeter Array). According to their results the age of most distant galaxy identified by JWST
is accurately established with Oxygen observation and is equal to 367 million years after the Big Bang.

ALMA also confirmed an existence of a hyper-luminous radio-loud AGN at
$z = 6.853$~\cite{endsley} in extremely massive reionization-era galaxy with 
$M_* = 1.7 \times  10^{11} M_\odot$. Such a very high AGN luminosity suggests that this object is powered by
 $ 1.6 \times 10^9 M_\odot$ black hole if accreting near the Eddington limit.
Nearly impossible, but PBH could seed such a monster.

In ref.~\cite{labbe} six candidate massive galaxies with
stellar mass $ > 10^{10} M_\odot$  at $7.4 <z <9.1$ i.e  500–700 Myr after the Big Bang have been found,
including one galaxy with a possible stellar mass of  $\sim 10^{11} M_\odot$, too massive
to be created in so early universe. As is stated in the publication, according to the ’science’ it is impossible to
create so well developed galaxies. The authors even suggest that they are supermassive black
holes of the kind never seen before. That might mean a revision of usual
understanding of black holes.  NB: This conclusion very well agrees with our predictions of PBHs.

\section{Seeding of cosmic objects by black holes \label{s-seeding}}

The hypothesis \cite{DS} (1993) and \cite{DKK} (2006), that SMBH seeded
galaxy formation allows to explain presence of SMBH in all large
and several small galaxies accessible to observation. This mechanism
explains how the galaxies observed by JWST in the very young universe, as well as in the present day universe,
might be created. 

Though the idea of seeding of the cosmic structures by "parent" black holes has been accepted in
several recent publications (see the following subsection), still the BH that performed the seeding are supposed to be created 
rather late in the process of the cosmic evolution. That creates in turn the problem of their creation. Much more natural 
idea, that the seeds are primordial black holes, is not accepted by the astrophysical community in spite of  serious observational 
evidence in favour of primordial black hole existence~\cite{pbh-evi}.

\subsection{Seeding in the early universe \label{ss-seed-eary}}

This idea of seeding was rediscovered in a few recent works. For example in ref.~\cite{bromm} is
stated that recent observations with
JWST have identified several bright galaxy candidates at $z \gtrsim 10$, some of
which appear unusually massive (up to $\sim 10^{11} M_\odot$). Such early formation
of massive galaxies is difficult to reconcile with standard $\Lambda$CDM
predictions, The observed massive galaxy candidates can be explained
if structure formation is accelerated by massive ($ \gtrsim 10^9 M_\odot$) PBHs
that enhance primordial density fluctuations.

As it is stated in ref.~\cite{bogdan-1}: detection of an X-ray quasar in a gravitationally-lensed z = 10.3
galaxy suggests that early supermassive black holes originate from heavy seeds.
The authors noted that "observations of high-redshift quasars reveal that many supermassive black holes
were in place less than 700 Million years after the Big Bang.
However, the origin of the first BHs remains a mystery.'' To resolve the puzzle a population of 
seeds of the first BHs are postulated. They are assumed to be either light i.e., $(10 - 100 )M_\odot$ 
being the remnants of the first stars or heavy i.e., $(10^4 - 10^5)M_\odot$, originating from direct
collapse of gas clouds. The latter  looks rather unnatural.
Much simpler and straightforward if the seeds are primordial BH, as predicted in refs.~\cite{DS,DKK}.

In the same paper~\cite{bogdan-1} the detection of an X-ray-luminous quasar powered by SMBH with the
mass $\sim 4 \times 10^7 M_\odot$ in the galaxy identified by JWST at $z \approx 10.3$ is reported.
It is argued that the detection of such quasar 
suggests that early supermassive black holes also originate from heavy seeds.
The mass of this QSO
is comparable to the inferred stellar mass of its host galaxy, in
contrast to the usual examples (but not always) from the local universe where mostly the
BH mass is $\sim  0.1\%$ of the host galaxy’s stellar mass. The combination of
such a high BH mass and large BH-to-galaxy stellar mass ratio $\sim 500$ Myrs
after the Big Bang is consistent with a picture wherein such BHs
originated from heavy seeds. 

This conclusions coincide with the predictions made in refs.~\cite{DS,DKK} but the mechanisms of seed creation are
drastically different.

Further investigation of the seeding problem applied to early SMBH observations is performed in ref. \cite{goulding}.
The James Webb Space Telescope detects early black holes (BHs) as they transform from seeds to supermassive BHs. 
In ref.~\cite{bogdan-1} the detection of an X-ray luminous supermassive black hole with a photometric redshift at z > 10 is reported.
According to the authors s uch an extreme source at this very high redshift provides new insights on seeding and
growth models for BHs given the short time available for formation and growth.
The resulting ratio of the black hole mass to the stellar mass remains two to three orders of magnitude
higher than local values, thus lending support to the heavy seeding channel for the formation of supermassive BHs within the first 
billion years of cosmic evolution.

\subsection{Seeding in contemporary universe \label{ss-seed-today}}

Primordial IMBHs with masses of a few thousand solar mass could
explain formation of globular clusters (GCs) and dwarf galaxies, as is predicted
in ref.~\cite{AD-KP}. In the last several years such IMBH
inside GSs and dwarfs are observed.

The seeding of dwarfs by intermediate mass BHs is confirmed by the recent
data, e.g. in the dwarf galaxy SDSS J1521+1404 the BH is discovered with
the mass $M \sim  10^5 M_\odot$ that is contains by far too much matter to be 
created by accretion inside such tiny creature.

The observation of two Candidates for Dual AGN in Dwarf-Dwarf Galaxy Mergers~\cite{dual-AGN}:
for the first time, astronomers have
spotted evidence of a pair of dwarf galaxies featuring giant black holes on
a collision course with each other. In fact, they haven’t just found just one
pair – they’ve found two.

In ref.~\cite{IMBH-dwarf}  a discovery of an
intermediate-mass black hole (IMBH) with a mass of
$M_{BH} = 3.6^{+5.9}_{-2.3} \times 10^5 M_\odot $, is reported, 
that surely cannot be created by accretion but might seed the dwarf formation.

To summarise, a large amount of observational data are at odds with the
conventional model but nicely fits the model of creation of primordial black holes
and primordial stars proposed in A. Dolgov, J.Silk, PRD 47 (1993) 4244, (DS)
and further developed in A.Dolgov, M. Kawasaki, N. Kevlishvili, Nucl. Phys. B807
(2009) 229, (DKK) The proposed mechanism is the first where inflation and
Affleck-Dine baryogenesis are applied to PBH formation, repeated now in a
number of works.
The impressive feature of the mechanism is the log-normal mass spectrum which
is the only known spectrum tested by experiment, in perfect agreement.


\section{Gravitational waves from BH binaries \label{s-GW-BH}}

GW discovery by LIGO strongly indicate that the sources of GW are PBHs, see e.g.~\cite{BDPP}.
There are several features that indicated to primordial origin of the GW sources:\\
1. Origin of heavy BHs, $M\sim 30M_\odot$, via conventional astrophysical processes is problematic,
though is not excluded.  Recently there appeared much more striking
problem of BH observation with $M \sim 100M-\odot$. To form so heavy BH, the progenitors should have 
$M > 100M_\odot$ and a low metal abundance to avoid too much mass loss during the evolution. Such heavy
stars might be present in young star-forming galaxies but they are not yet observed. A possible way out is
suggested in ref.~\cite{Zieg-KF} by suggestion that possible dark matter annihilation inside stars could
permit formation of BHs with $M \gtrsim 100M_\odot$. \\
On the other hand, PBHs with the observed by LIGO masses may be easily created with sufficient density.

2.  Formation of BH binaries from the original stellar binaries has quite low probability.
Stellar binaries are formed from common interstellar gas clouds and are quite frequent in galaxies. If BH is
created through stellar collapse, a small non-sphericity results in a
huge velocity of the BH and the binary is destroyed. BH formation
from PopIII stars and subsequent formation of BH binaries with tens
of M is estimated to be small. The problem of the binary formation
is simply solved if the observed sources of GWs are the binaries of
primordial black holes. They were at rest in the comoving volume,
when inside horizon they are gravitationally attracted and may loose
energy due to dynamical friction in the early universe. The probability
to become gravitationally bound is quite significant.\\
The conventional scenario is not excluded but  looks much less natural

3. The low values of the BH spins  in almost all except for
three events strongly constrains astrophysical BH formation from close
binary systems. Astrophysical BHs are expected to have considerable angular
momentum, nevertheless the dynamical formation of double massive low-spin BHs
in dense stellar clusters is not excluded, though difficult. On the other hand, PBH
practically do not rotate because vorticity perturbations in the early universe are
vanishingly small.
However, individual PBH forming a binary initially rotating on elliptic orbit could
gain collinear spins about 0.1 - 0.3, rising with the PBH masses and
eccentricity~\cite{post-mit,post-kur-mit}. This result is in agreement with the GW170729 LIGO event
produced by the binary with masses 50and 30M and GW190521.

\subsection{Chirp mass distribution  \label{ss-chirp}}

Two rotating gravitationally bound massive bodies are known to emit
gravitational waves. In quasi-stationary inspiral regime, the radius of the orbit and
the rotation frequency are approximately constant and the GW frequency is twice
the rotation frequency. The luminosity of the GW radiation is:
\be 
L = \frac{32}{5}\,m_{Pl}^2 \left(\frac{M_c \omega_{orb} }{M_{Pl}^2}\right)^{10/3},
\label{Lbin}
\ee
where $m_{Pl} = 1.22\times 10^{19} $ GeV is the Planck mass and
$M_c$ is the so called chirp mass:
\be
M_c = \frac{(M_1 M_2)^{3/5}}{M_1 + M2)^{1/5}}
\label{M_c}
\ee
and $M_1$ and $M_2$ are the masses of two bodies in the binary system. 
The orbital frequency is equal to:
\be
\omega_{orb}^2 = \frac{M_1 + M_2}{m_{Pl}^2 R^3},
\label{omega-orb}
\ee
where $R$ is the radius of the orbit.

In ref.~\cite{DKMPPSS} he available data on the chirp mass distribution of the black holes in the
coalescing binaries in O1-O3 LIGO/Virgo runs are analyzed and compared
with theoretical expectations based on the hypothesis that these black
holes are primordial with log-normal mass spectrum.
The inferred best-fit mass spectrum parameters, $M_0 = 17M_\odot$ and
$\gamma = 0.9$, fall within the theoretically expected range and shows excellent
agreement with observations, see fig. 4 .

\begin{figure}[htbp]
\begin{center}
\includegraphics[scale=0.13]{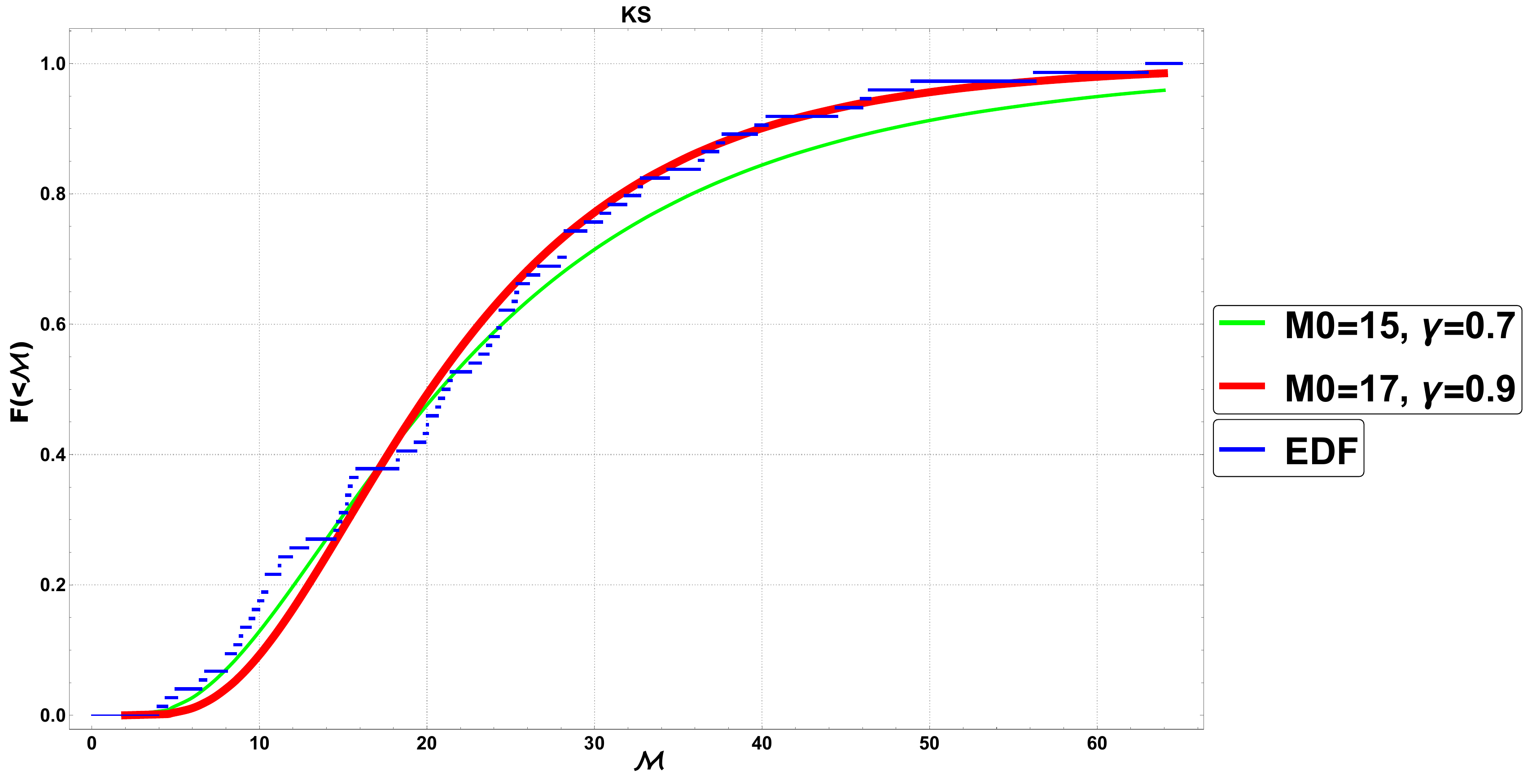}
\caption{Model distribution $F_{PBH}(< M)$ with parameters  $M_0$ and $\bm{\gamma}$ for two best 
Kolmogorov-Smirnov tests.  EDF= empirical distribution function.}
\end{center}
\end{figure}

In Fig. 5 cumulative distributions $F(< M)$ for several { astrophysical} models of binary BH coalescences are presented. 
Evidently binary black hole formation based on massive binary
star evolution requires additional adjustments to reproduce the observed
chirp mass distribution

\begin{figure}[htbp]
\begin{center}
\includegraphics[scale=0.15]{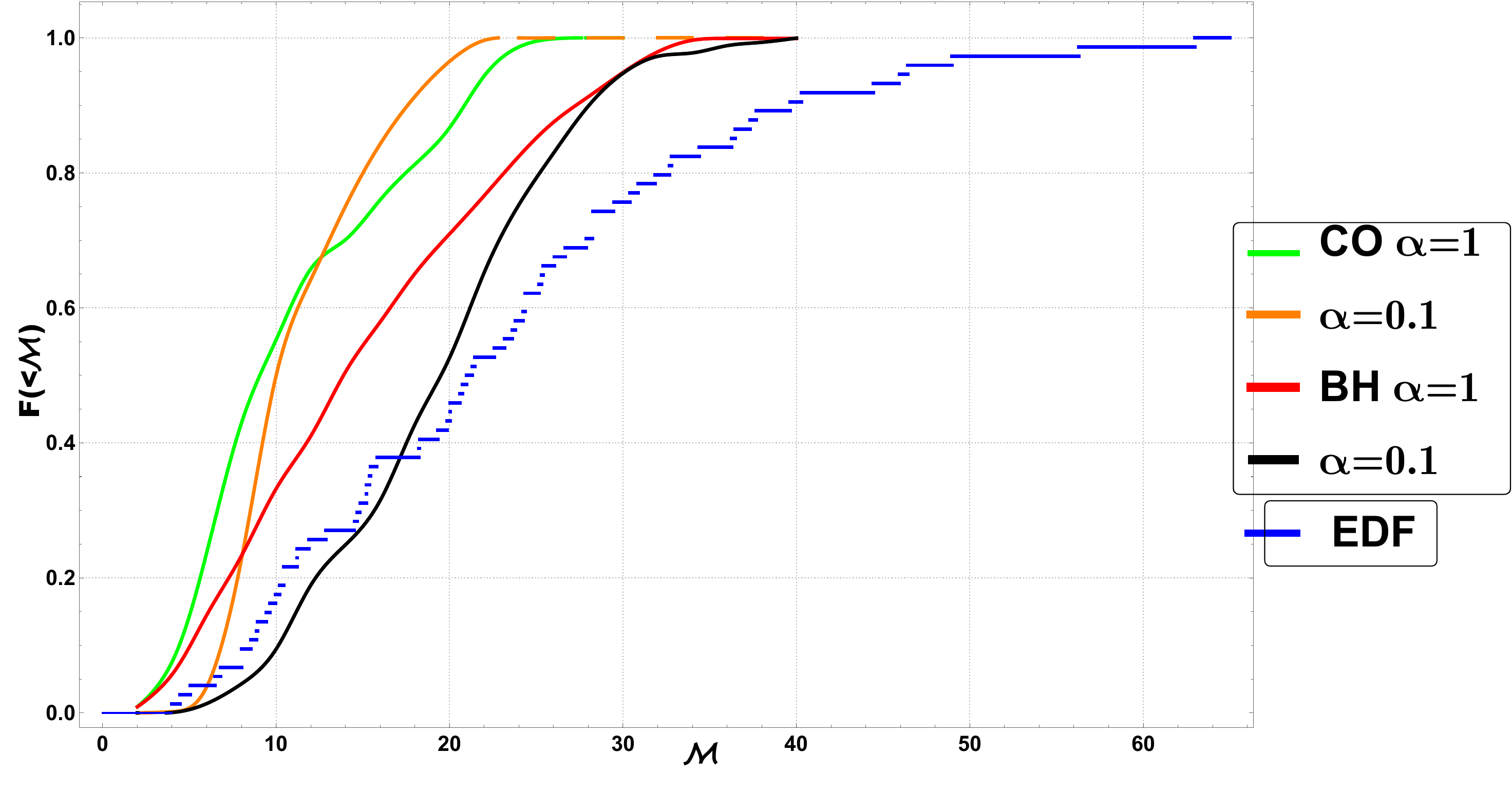}
\caption{Chirp mass distribution of astrophysical BHs compared to observations (blue)}
\end{center}
\end{figure}

To conclude, PBHs with log-normal mass spectrum perfectly fit the data, while
astrophysical BHs seem to be disfavoured.

An analysis of the new set of Ligo-Virgo-Kagra data was performed recently in ref.~\cite{Pos-Mit-new}.
The authors dfound that the chirp-mass distribution of LVK GWTC-3 BH+BH binaries has two distinct bumps that could be
explained by two different populations of BH+BH binaries:
1) the low-mass bump with $M_0 \sim 10M_\odot$ due to the astrophysical BH+BH formed
in the local Universe from the evolution of massive binaries;\\
2) the PBH binaries with log-normal mass spectrum with $M_0 \sim 10M_\odot$ and $\gamma \sim 10$. 
The central mass of the PBH distribution is somewhat larger than the expected
PBH mass at the QCD phase transition, $M_0 \sim  8M_\odot$ but still can be accommodated
with the mass of the cosmological horizon provided that the temperature of the
QCD phase transition $T_{PT} \sim 70$ MeV, that is possible for a large chemical potential. However, 
the huge value of $\gamma$,  corresponding to the second bump, leads to a sharp BH mass distribution sharply
concentrated  near $M=M_0$. 

\section{Black dark matter \label{s-black-DM}} 

The first suggestion PBH might be dark matter ”particles” was made by S.
Hawking in 1971~\cite{SH-dm} 
and later repeated by G. Chapline in 1975~\cite{chaplin} who assumed the flat mass
spectrum in log interval: $dN = N_0(dM/M)$
with maximum mass $M_{max} \approx10^{22}$ g, which hits the allowed mass range,
according to the review~\cite{Carr-Kuh-1,Carr-Kuh-1}. 
The next work~\cite{DS} of 1993, entitled  ''Baryon isocurvature fluctuations
at small scales and {\bf baryonic dark matter}", a more interesting mass range with much higher BH masses was proposed,
in particular with masses observed by LVC (LIGO-Virgo-Kagra).

Large values of BH masses strongly exceeding the horizon mass in the early universe was allowed, because of the 
pioneering suggestion that cosmological inflation was invoked for the PBH formation. As a result the PBHs with
masses as high as $10^6 M_\odot$ and even higher can be created.

Constraints on PBHs on the cosmological fraction of the mass density as a function of mass are reviewed  in 
refs.~\cite{Carr-Kuh-1,Carr-Kuh-2} and presented in Fig. 6.
The  bounds are derived for the monochromatic mass spectrum of BHs and according to the
authors are model dependent and have caveats. 
\begin{figure}[htbp]
\begin{center}
\includegraphics[scale=0.5]{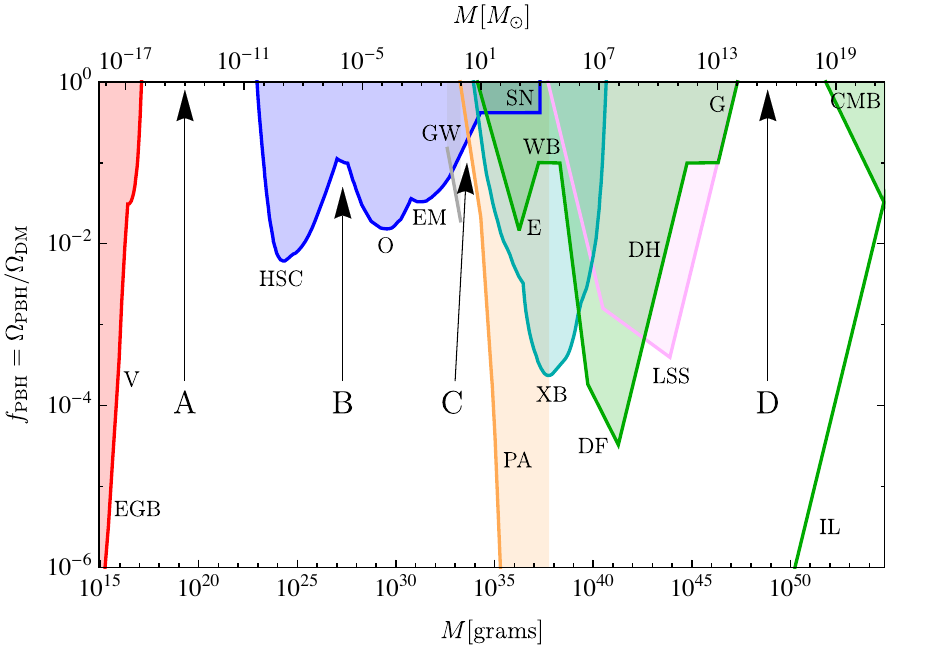}
\caption{Allowed fraction of the cosmological BH density}
\end{center}
\end{figure}

As is argued in ref.~\cite{rubin} PBHs can be formed in clusters.
Dynamical interactions in PBH clusters offers additional channel for the orbital
energy dissipation thus increasing the merging rate of PBH binaries, and the
constraints on fPBH obtained by assuming a homogeneous PBH space distribution
can be weaker. A recent analysis~\cite{er-stas}  based on the PBH formation model \cite{sasaki} and \cite{nakamura}
shows that the cosmological mass fraction even up to $f_{PBH} = 1$ is not excluded \footnote{I thank K. Postnov for indicating
the above references.}

\section{Antimatter in cosmology and in the Galaxy \label{s-antimatter}}

Astronomical data of the several recent years present strong evidence in favour of
noticeable antimatter population in our Galaxy including:\\
• Observation of gamma-rays with energy 0.511 MeV, which surely originate from
electron-positron annihilation at rest.\\
• Very large flux of anti-helium nuclei, observed at AMS.\\
• Several stars are found which produce excessive gamma-rays with
energies of several hundred MeV which may be interpreted as indication
that these stars consist of antimatter.

\subsection{Antimatter history \label{ss-anti-history}}

The search for galactic antimatter (anticomets) was pioneered by 
B.P. Konstantinov and collaborators~\cite{konst-1,konst-2} in 1968 and 1969.

Antimatter in the universe was first considered by F.W. Stecker starting for 1971~\cite{steck-1,steck-2}
Summary of the situation at 2002 was presented at two lectures~\cite{FS,AD}.

The father of antimatter is justly considered Paul A.M. Dirac who in his Nobel lecture in 1933 
“Theory of electrons and positrons”, anticipated
”It is quite possible that... these stars being built up
mainly of positrons and negative protons. In fact, there may be half the
stars of each kind. The two kinds of stars would both show exactly the
same spectra, and there would be no way of distinguishing them by present
astronomical methods.” It is shown in ref.~cite{DNV}
that this is not exactly so.  As it was srgued in ref.~\cite{DNV}
the spectra are not exactly the same, even if CPT is unbroken and the
polarisation of radiation form weak decays could be a good indicator or the
type of emitted neutrinos/antineutrinos from (anti)supernovae. However, observations 
of the effects indicated in~\cite{DNV} are not realistic at the present days.

In fact Dirac was the second person to talk about antimatter. In 1898, 30
years before Dirac and one year after discovery of electron (J.J. Thomson,
1897) Arthur Schuster (another British physicist) conjectured that there
might be other sign electricity, {\it antimatter}, and supposed that there
might be entire solar systems, made of antimatter, indistinguishble
from ours. Schuster’s wild guess: matter and antimatter are capable to annihilate and
produce vast energy surprisingly happened to be true. On the other hand
he believed that matter and antimatter  were gravitationally repulsive since antimatter had negative mass.
Two such objects on close contact should have vanishing mass!?
Quoting his paper\cite{Schuster}: “When the year’s work is over and all sense of responsibility has left us,
who has not occasionally set his fancy free to dream about the unknown,
perhaps the unknowable?... Astronomy, the oldest and yet most juvenile of the sciences, may still have
some surprises in store. May antimatter be commended to its case".

\section{Antimatter in the Milky Way \label{s-antiGAL}}

Based on the conventional approach no antimatter object is expected to
be in the Galaxy. However, it was predicted in 1993~\cite{DS} and  
elaborated in 2009!\cite{DKK} that noticeable amount
of antimatter, even antistars might be in the Galaxy and in its halo.

Bounds on the density of galactic antimatter and in particular on
antistars are quite loose~\cite{CB-AD,DB,BDP}. The reason for that
is a relatively small energy release in the process of annihilation because it
takes place only in a thin surface layer of the star.

\subsection{Galactic positrons \label{ss-positrons}}

Observation of intense 0.511 line is a proof of abundant positron population in the
Galaxy. In the central region of the Galaxy electron–positron annihilation proceeds
at a surprisingly high rate, creating the flux:
\be
\Phi_{511 \; {\rm keV}} = {1.07 \pm 0.03 \cdot 10^{-3} }\; 
{\rm photons \; cm^{-2} \, s^{-1}} .
\label{Phi}
\ee

The width of the line is about 3 keV. Emission mostly goes from the Galactic
bulge and at much lower level from the disk~\cite{e+1,e+2,e+3}
Until recently the commonly accepted explanation was that $e^+$ are
created in the strong magnetic fields of pulsars but the recent results of
AMS probably exclude this mechanism, since the spectrum of $\bar p$  and $e^+$ at
high energies are identical~\cite{Ting-2021}. 

\subsection{Galactic antinuclei \label{ss-antinuc}}

In 2018 AMS-02 announced possible observation of six
${\bar{He}^3}$ and two ${\bar{He}^4}$~\cite{AMS-antinuc}.

In 2022 more events have been presented~\cite{AMS-antinuc-2}:
7 $\bar D$ ($E \lesssim 15$ GeV) and 9 $\overline{He}$ ($E \sim 50$ GeV).
The ratio of the fluxes ${\overline{He}/He \sim 10^{-9}}$ is too high to be explained by the secondary
production of antinuclei in by cosmic rays.   
It is not excluded that the flux of anti-helium is even much higher because low energy 
${\overline{He}}$ may escape registration at AMS.

Antinuclei creation in cosmic rays are estimated in ref.~\cite{anti-nuc-CR}
Anti-deuterium can be created in the collisions
${\bar p\,p}$ or ${\bar p\, He}$ that would produce the flux of $\bar D$.
${\sim 10^{-7} /m^{2}/ s^{-1}} $/steradian/GeV/neutron), i.e. 5 orders of magnitude below the observed flux of antiprotons.

The fluxes of   ${\overline{He}^3}$ and ${\overline{He}^4}$, that could be created in 
cosmic rays are respectively 4 and 8
orders of magnitude smaller than the flux of anti-D.

After the AMS announcement of observations of  anti-$He^4$ there appeared several
theoretical attempts to create  anti-$He^4$ through dark matter annihilation. It looks quite unnatural, especially
because the main channel of annihilation is expected to be elementary particles but not bound systems of antiprotons
and antineutrons.

{There is noticeable discrepancy between the large fraction of D with respect to He. 
In the case of the standard BBN this ratio
should be smaller than unity, but the observed one is practically 1.}
 
According to the model~\cite{DS,DKK}
the abundances of D and He are determined by BBN with large baryon-to-photon ratio
$\beta$ (or $\eta$). However if $\beta \sim 1$ there is no primordial D. On the other hand in our scenario  formation of primordial 
elements takes place inside non-expanding compact stellar-like objects with fixed temperature. If the temperature is sufficiently 
high, this so called BBN may stop before abundant He formation 
with almost equal abundances of D and He. One can see that looking at 
abundances of light elements as a function of temperature. 
{If it is so, antistars may have equal amount of 
$\overline{D}$ and $\overline{He}$.}
 
\subsection{ Antistars in the Galaxy. \label{ss-antistars}} 

In ref.~\cite{gal-anti-stars} surprising evidence for antistar observation in the Galaxy are presented.
Quoting the authors:
 "We identify in the catalog 14 antistar candidates not associated with any objects belonging 
to established gamma-ray source classes and with a spectrum compatible with 
baryon-antibaryon annihilation.'' The positions of these possible antistars are presented in Fig.~7.
\begin{figure}
\includegraphics[scale=0.45]{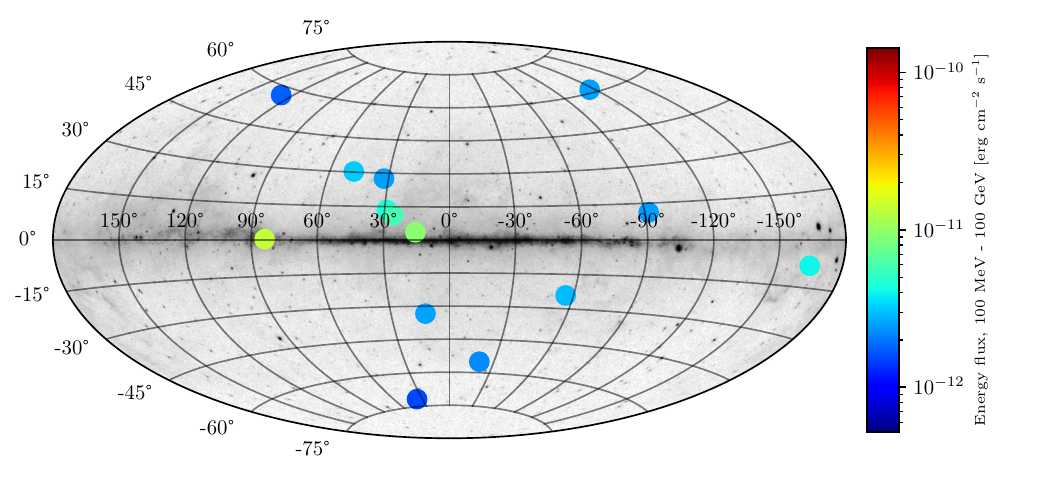}
\caption{\label{fig:sources} Positions and energy flux in the 100 MeV - 100 GeV range of antistar candidates selected in 4FGL-DR2. Galactic coordinates. The background image shows the Fermi 5-year all-sky photon counts above 1 GeV }
\end{figure}  

An alternative way to identify antistars is proposed in paper~\cite{BBBDP}.
In astrophysically plausible cases of the interaction of neutral atmospheres or winds from 
antistars with ionised interstellar gas, the hadronic annihilation 
{ will be preceded by the formation of excited $ {p \bar p}$
and $He {\bar p}$ atoms.} These atoms rapidly cascade down to low levels prior to 
annihilation giving rise to a series of narrow lines which can be associated with the hadronic 
annihilation gamma-ray emission. The most significant are L (3p-2p) 1.73 keV line (yield more 
than 90\%) from ${p \bar p}$ atoms, and M (4-3) 4.86 keV (yield $\sim 60$\%) and L (3-2) 11.13 
keV (yield about 25\%) lines from $He^4 \bar p$ atoms. These lines can be probed in dedicated 
observations by forthcoming sensitive X-ray spectroscopic missions XRISM and Athena and in 
wide-field X-ray surveys like SRG/eROSITA all-sky survey.

Possible sources of antinuclei in cosmic rays from antistars which are
predicted in a modified Affleck-Dine baryogenesis scenario of~\cite{DS,DKK}
are studied in ref.~\cite{BPBBD}. 
The expected fluxes and isotopic content of antinuclei in the
GeV cosmic rays produced in scenarios involving antistars are estimated.
It is shown that the flux of antihelium cosmic rays reported by the AMS-02
experiment can be explained by Galactic anti-nova outbursts,
thermonuclear anti-SN Ia explosions, a collection of flaring antistars, or an
extragalactic source with abundances not violating existing gamma-ray and
microlensing constraints on the antistar population.

\section{Anti-creation mechanism \label{s-anti-creation}}

The model of creation of antimatter in the galaxy is a byproduct of the~\cite{DS,DKK} scenario
of  creation of massive primordial black holes. It is based on supersymmetry (SUSY)
motivated baryogenesis, proposed by Affleck and Dine (AD)~\cite{aff-dine}.  
SUSY predicts existence of  bosons with non-zero baryonic number ${ B\neq 0}$.
Such bosons may condense along flat directions of the quartic potential:
\be{{
U_\lambda(\chi) = \lambda |\chi|^4 \left( 1- \cos 4\theta \right)
}}
\ee
and of the mass term, ${ U_m=m^2 \chi^2 + m^{*\,2}\chi^{*\,2}}$:
\be
U_m( \chi ) = m^2 |\chi |^2  \left[ 1-\cos (2\theta+2\alpha) \right],
\ee
where ${ \chi = |\chi | \exp (i\theta)}$ and ${ m=|m|e^\alpha}$.
In AD scenatio baryonic number is naturally non-conserved .

After inflation ${\chi}$ was away from the origin and, when 
inflation was over, started to evolve down to equilibrium point, ${\chi =0}$,
according to the equation:
\be{{
\ddot \chi +3H\dot \chi +U' (\chi) = 0.
}}
\ee
Baryonic charge of $ \chi$,
$B_\chi =\dot\theta |\chi|^2$,
is analogous to mechanical angular momentum in complex $\chi$ plane. 
${{\chi}}$ decays transferred
baryonic charge to that of quarks in B-conserving process.

{AD baryogenesis could lead to baryon asymmetry of order of unity, much larger
than the observed  ${10^{-9}}$.}



If $ { m\neq 0}$, 
the angular momentum, B, is generated by different 
directions of the  quartic and quadratic valleys at low ${\chi}$.
{If CP-odd phase ${\alpha}$ is small but non-vanishing, both baryonic and 
antibaryonic domains might be  formed}
{with possible dominance of one of them.}
 
 A new input of the model~\cite{DS,DKK} is the coupling of $ \chi$ 
 to inflaton $ \Phi$ (the first term in the equation below):
\be 
U = {g|\chi|^2 (\Phi -\Phi_1)^2}  +
\lambda |\chi|^4 \,\ln \left( \frac{|\chi|^2 }{\sigma^2 } \right)
\nonumber\\
{+\lambda_1 (\chi^4 + h.c. ) + 
(m^2 \chi^2 + h.c.). \,\,\,\,\,\,\,\,\,\,\,\,\,\,\,\,\,\,\,\,
}
\ee
{Coupling to inflaton is a general renormalizable one.}
{When the window to the flat direction is open, near ${\Phi = \Phi_1}$, }
{the field ${\chi}$ slowly diffuses to large value,} according to quantum diffusion
equation derived by Starobinsky, generalised to a complex field $\chi$.

If the window to flat direction, when ${\Phi \approx \Phi_1}$ is open only {during 
a short period,} cosmologically small but possibly astronomically large 
bubbles with high ${ \beta}$ could be
created, occupying {a small
fraction of the universe,} while the rest of the universe has normal
{${{ \beta \approx 6\cdot 10^{-10}}}$, created 
by a small ${\chi}$}. 
{This mechanism of massive PBH formation quite different from all others.}
{The fundament of PBH creation is build at inflation by making large isocurvature
fluctuations at relatively small scales, with practically vanishing density perturbations.} 
Initial isocurvature perturbations are in chemical content of massless quarks.
Density perturbations are generated after the QCD phase transition.
{The emerging universe looks like a piece of Swiss cheese, where holes are high baryonic 
density objects occupying a minor fraction of the universe volume.}\\
 {\bf Results of (Anti-)Creation.}\\
 1.PBHs with log-normal mass spectrum - confirmed by the data!\\
2. {Compact stellar-like objects, as e.g. cores of red giants.}\\
3. {Disperse hydrogen and helium clouds  with (much) higher than average $\bm{n_B}$ density.} Strange stars with unusual
chemistry and velocity.\\
4.{${\beta}$ may be negative leading to creation of
(compact?) antistars which could survive annihilation with the 
homogeneous baryonic background.}\\
5.Extremely old stars would exist.  Several such stars are
observed. Even, "older than universe star"  is found~\cite{older-star}; the older age is mimicked by the unusual initial chemistry.


  \section*{FUNDING}
This work was supported by RSF Grant 23-42-00066.


\nocite{*}


\end{document}